\documentclass{ws-procs9x6}
\usepackage{makeidx}
\makeindex
\begin{document}
\title{KINETICS OF VACUUM PAIR CREATION IN STRONG ELECTROMAGNETIC FIELDS
\footnote{Based on a poster presented at the conference ``Progress
in Nonequilibrium Greens Functions, Dresden, Germany, 19.-22.
August 2002''}}
\author{A.V. Tarakanov, A.V. Reichel and S.A. Smolyansky}
\address{\it Physics Department, Saratov State University, 410071, Saratov, Russia\\
 E-mail: smol@sgu.ru}
\author{D.V. Vinnik}
\address{\it Institute of Theoretical Physics, Tuebingen University, auf der Morgenstelle, \\
 14, 72076 Tuebingen, Germany\\
E-mail: vinnik@alpha11.tphys.physik.uni-tuebingen.de}
\author{S.M. Schmidt}
\address{\it Institute of Theoretical Physics, Tuebingen University, auf der Morgenstelle, \\
 14, 72076 Tuebingen, Germany\\
 Helmholtz-Gemeinschaft, Ahrstrasse 45, D-53175 Bonn, Germany\\
 E-mail: sebastian.schmidt@helmholtz.de}
\maketitle
\abstracts{We study particle - antiparticle pair
production\index{Particle - antiparticle pair production} under
action of a strong time dependent space homogeneous electric field
at the presence of a collinear constant magnetic
field\index{Strong electromagnetic fields}. We derive the kinetic
equation\index{Kinetic equation} for a such  field configuration
for fermions and bosons in the framework of the Schwinger
mechanism\index{Schwinger mechanism} of vacuum tunneling.  We show
the enhancement of pair production  for fermions (suppression for
bosons) with the increasing of the magnetic field as in the case
of the constant electromagnetic field. We have constructed closed
set of equations, which can be applied to some actual problems
with manifestation of  strong electromagnetic fields\index{Strong
electromagnetic fields}, e.g., it is essential in the framework of
the Flux Tube Model of Quark - Gluon Plasma generation; for
describing some cosmological objects and especially because of the
planned experiments on creation of subcritical fields in the
X-Free Electron Laser pulses.}

\section{Introduction}
It is expected that the influence of  a magnetic field on vacuum
particle production at the presence of a non-stationary electric
field can be essential in many physical situations. It should be
mentioned, first of all, that the joint consideration of
chromoelectric and chromomagnetic fields is necessary while
constructing the Flux Tube Model(FTM) of superconductive type when
describing the pre-equilibrium evolution of Quark - Gluon
Plasma(QGP), generated under extreme conditions of
ultra-relativistic heavy ion collisions \cite{Lam,CRSS}. It is expected
that extra-strong electromagnetic fields\index{Strong
electromagnetic fields} can exist in a series of astrophysical
(e.g., magnetars)\index{magnetars} and cosmological (e.g., cosmic
strings) objects. Finally, when achieving impact near-critical
magnetic fields in a laboratory environment in comparatively short
distances it would be very perspective to study combined action of
electric and magnetic fields on vacuum $e^+e^-$ pair
creation\index{Electron-positron pair production}, e.g., under
conditions of the planned experiment on the X-Free Electron
Laser(X-FEL) \cite{vsp}.

Electron-positron pair production\index{Electron-positron pair
production} in constant electromagnetic(EM) fields was
considered in the work \cite{Niki} (see also \cite{Grib}). Here we
will assume the electric field is time dependent (but the magnetic
field remains constant). For derivation of the kinetic equation
(KE)\index{Kinetic equation} we will use non-perturbative approach
developed in works\cite{We}.

\section{Solution of the one-particle problem}
We consider $e^+e^-$\index{Electron-positron pair
production} vacuum pair creation under action of the external
EM field with the configuration of vector potential
of the kind
\begin{equation}\label{eq1}
  A^{\mu}(\vec{x},t) = \left(0 ,0, -Hx^1, A^3(t) = A(t)\right)\,,
\end{equation}
where $H$ is the strength of the magnetic field, so that we have
space homogeneous EM field, polarized in one
direction. The electric field is time dependent, whereas the
magnetic field is constant,
\begin{equation}\label{eq2}
\vec{E}(t) = \left(0 ,0, E^3(t) =- \dot A^3(t)\right), \qquad  \vec{H} = (0, 0, H)\,,
\end{equation}
where dot denotes time derivative.

The squared Dirac Equation is
$ \left(D^2 + m^2 + \frac{ie}{2}F_{\mu\nu}\gamma^{\mu}\gamma^{\nu}\right)\Psi = 0,$
where $ F_{\mu\nu} = A_{\nu,\mu} - A_{\mu,\nu}$, $D_{\mu} =\partial_{\mu} - ieA_{\mu}$ and
$-e$ is the electron charge.
This equation in the chosen configuration of fields (\ref{eq1}) has the form
\begin{eqnarray}\label{eq3}
(\partial^2_0 - \triangle + m^2 + 2ie[Hx^1\partial_2 - A\partial_3]
+ e^2 [H^2(x^1)^2 + A^2] +  \nonumber \\
+ie[\partial_0A\gamma^0\gamma^3 - H\gamma^1\gamma^2] ) \Psi = 0 \,.
\end{eqnarray}
Let us choose solution in the form
\begin{equation}\label{eq4}
  \Psi_r(\vec{x}, t) =  CT(t)\Psi(x_1)e^{i(p^2x^2 + p^3x^3)}R_r\,,
\end{equation}
where $C$ is a normalizing constant and $R_r (r=1,2)$ are
Nikishov spinors~\cite{Niki,Grib}:
\begin{equation}\label{eq5}
  R_1^{+} = (0, 1, 0, -1), \qquad  R_2^{+} = (1, 0, 1, 0) \,.
\end{equation}
The result of separation of variables gives the time-dependent part
\begin{equation}\label{eq6}
 \left\{\partial^2_0 + \omega^2_{r\lambda}(\vec{p}, t) + ie\partial_0A \right\}
  T_{\lambda r}(t) = 0\,,
\end{equation}
and space-dependent part
\begin{equation}\label{eq7}
\left\{\partial^2_1 + 2ex^1p^2H - e^2(x^1)^2H^2 + \lambda \right\}\Psi(x^1) = 0\,,
\end{equation} where $\lambda$ is a separation constant and
$$\omega^2_{r\lambda} (\vec{p}, t) = (p^2)^2 + (p^3 + eA)^2 + m^2 - e(-1)^rH + \lambda \,.$$
Eq. (\ref{eq7}) can be reduced to
\begin{equation}\label{eq8}
\Psi''_n(\eta) + (2n + 1 - \eta^2)\Psi_n(\eta) = 0, \, \, \mbox{with}\, \,
\eta =\sqrt{eH}\left(x^1 - \frac{p^2}{eH}\right)\,,
\end{equation}
where prime means derivative with respect to $\eta$.

The solution of Eq. (\ref{eq8}) is known
\begin{equation}\label{eq9}
 \Psi_n(\eta) = exp(-\eta^2/2) H_n(\eta), \,\, \mbox{with}\,\, \lambda_n = (2n + 1) eH \,,
\end{equation}
where $H_{n}$ are   Chebyshev-Hermite polynomials and we have
solutions
\begin{equation}\label{eq11}
  \Psi^{(\pm)}_{np^2p^3,r}(\vec{x}, t) = C_nT^{(\pm)}_{nr}(t)
  exp(-\eta^2/2)H_n(\eta)e^{i(p^2x^2 + p^3x^3)}R_r\,,
\end{equation}
where $(\pm)$ signs correspond to positive and negative frequency
solutions of the equation (\ref{eq6}) and we have introduced the
kinetic momentum $P = p^3+ eA(t)$ and one particle energy
$\omega^2_{rn}(P) = P^2 + m^2 + eH [2n+1+ (-1)^r]$.

Finally, the normalization conditions are
\begin{equation}\label{eq14}
\int d^3x \bar{\Psi}^{(\pm)}_{np^2Pr} (\vec{x}, t) \Psi^{(\pm)}_{n'p'^2P'r'} (\vec{x},t) =
\delta_{nn'}\delta_{rr'}\delta(p^2-p'^2)\delta(P-P')\,,
\end{equation}
with normalization constant
$C_n =  \left( \frac{eH}{\pi} \right)^{1/4} \frac{1}{2\pi} (2^{n+1}n!)^{-1/2}$.

Functions (\ref{eq11}) form the complete system of orthonormalized
functions. That is enough for the construction of the secondary
quantized representation with the field operator
\begin{equation}\label{eq15}
  \Psi(x) = \sum_{r,n} \int dP \left\{ \Psi^{(-)}_{np^2Pr} (\vec{x}, t) a^{(-)}_{np^2Pr}
  + \Psi^{(+)}_{np^2Pr} (\vec{x}, t) a^{(+)}_{np^2Pr} \right\}\,,
\end{equation} where $a^{(-)}, a^{*(+)}$ are annihilation and
creation operators with the standard set of anti-commutation
relations.

\section{ The kinetic equation in the quasi-particle representation}

The Kinetic Equation(KE)\index{Kinetic equation} derivation
requires the transition to the quasi-particle representation,
which is archived by the diagonalization procedure of the
Hamiltonian. In the considered case the procedure is distinguished
from the usual one \cite{Grib} only in some details, concerned
with the specific character of  basis functions (\ref{eq11}). Let
the new operators $b^{(-)}(t)$ and $b^{*(+)}(t)$ are annihilation
and creation operators in the quasi-particle representation,
connected with $a^{(-)}$ and $a^{*(+)}$ operators by the time
dependent Bogoliubov transformation \cite{Grib}. Then it is
possible to introduce the distribution function of electrons in
the new representation with the momentum $p^2, P$ and spin $r$ on
the Landau $n$- level
\begin{equation}\label{eq16}
f_{nr} (p^2, P, t) = \langle0_{in}|b^{*(+)}_{np^2Pr}(t)
b^{(-)}_{np^2Pr}(t)|0_{in}\rangle
\end{equation} and the corresponding distribution function of positrons
$\bar{f}_{nr}(p^2,P,t)$ with an electric neutrality condition $f=
\bar{f}$.

The KE\index{Kinetic equation} derivation  for function
(\ref{eq16}) differs from the case of $H=0$ \cite{We} only in some
details concerned with the source term describing the processes of
vacuum tunneling. Now we have
\begin{equation}\label{eq17}
  \dot{f}_{nr} (P,t) = S_{nr}(P,t)\,,
 \end{equation}
where source of pair production is
\begin{eqnarray}\label{eq18}
\begin{array}{lcl}
 &&S_{nr}(p^2, P, t) = S_{nr}(P,t)= \frac{1}{2} w_{nr}(P, t) \times \\ \\
 &\times& \int^t_{-\infty}  dt'w_{nr}(P(t,t'), t') [1-2f_{nr}(P,t')]
  \cos{ \left\{ 2\int^t_{t'} d\tau \omega_{nr}(P,\tau)  \right\}} \,,
\end{array}
\end{eqnarray}
where $P(t,t') = P - e \int^t_{t'} d\tau E(t)\,$ and
\begin{equation}\label{eq19}
  w_{nr}(P, t) =  \frac{eE(t)\varepsilon_{nr}}{\omega^2_{nr}(P)}\,,
\end{equation}
where $\varepsilon^2_{nr} = m^2 + eH[1+2n+(-1)^r]$.

As one can see from Eqs. (\ref{eq18})-(\ref{eq19}) the
distribution function does not depend on $p^2$, $ f_{nr} (p^2,P,t)
= f_{nr} (P,t)$ (the cylindrical symmetry of the considered
problem). This circumstance was taken into account in the
KE\index{Kinetic equation} Eq.~(\ref{eq17}).

The comparison of the source term (\ref{eq18})-(\ref{eq19}) with
the "old" case $H=0$~\cite{We} shows that the "new" quasi-particle
frequency $\omega_{rn}$, transition amplitude (\ref{eq19}) and
transversal energy $\varepsilon_{nr}$ can be obtained from the
corresponding "old" formulas by means of formal substitution
\begin{equation}\label{eq23}
  (p^1)^2+(p^2)^2=p^2_{\bot} \rightarrow eH[1+2n+(-1)^r]\,,
\end{equation}
which provides discrete transversal energy as in the case with boundary
conditions~\cite{FBS}

For numerical analysis it is convenient to reduce the
KE\index{Kinetic equation} (\ref{eq17})-(\ref{eq19}) to the
following system of ordinary differential equations:
 \begin{eqnarray}\label{eq24}
  \dot{f}_{nr}(P,t) &=& \frac{1}{2}w_{nr}(P,t) v_{nr}(P,t)\,, \nonumber \\
  \dot{v}_{nr}(P,t) &=& w_{nr}(P,t)(1-2f_{nr}(P,t))-2\omega_{nr}(P,t) z_{nr}(P,t)\,, \\
  \dot{z}_{nr}(P,t) &=& 2 \omega_{nr}(P,t) v_{nr}(P,t)\,, \nonumber
\end{eqnarray}
where $v$ and $z$ are auxiliary functions.

\section{Scalar particle production}

The Klein- Gordon equation in the field (\ref{eq1}) permits the
separation of variables as well
\begin{equation}\label{eqs1}
  \varphi^{\pm}(x)=T^{\pm}(t)\Phi(x_{1})e^{ip^{2}x^{2}+ip^{3}x^{3}}\,,
\end{equation}
where functions $T(t)$ and $\Phi(x_{1})$ satisfy following equations
\begin{eqnarray}\label{eqs2}
\ddot{T}^{\pm}+\omega^{2}_{\lambda}(t)T^{\pm}=0\,, \qquad
\label{eqs3}
\Phi''(\eta)-\eta^{2}\Phi(\eta)+\frac{\lambda+(p^{2})^{2}}{|e|H}\Phi(\eta)=0\,,
\end{eqnarray}
where now $\eta=\sqrt{|e|H}\left(x^{1}+p^{2}/|e|H\right)$.
Normalized solutions of the last equation are again
Chebyshev-Hermite polynomials with
$\lambda_{n}+(p^{2})^{2}=(2n+1)|e|H$ , $n=0,1,\ldots$. Functions
$T^{\pm}(t)$ are positive and negative solutions of (\ref{eqs2})
and we can write a formal solution
\begin{equation}\label{eqs4}
  \varphi^{\pm}_{n,p^{2},p^{3}}(x)=C_{n}T^{\pm}_{n}(t)exp(-\eta^{2}/2)H_{n}(\eta)e^{i(p^{2}x^{2}+p^{3}x^{3})},
\end{equation}
where a normalization constant is
$C_{n}=\frac{1}{2\pi}\left(\frac{|e|H}{\pi}\right)^{1/4}\left(2^{n}n!\right)^{-1/2}$, which
can be obtained from the condition
$$i\int d^{3}x\varphi^{(\pm)*}_{n,p^{2},p^{3}}(x)\stackrel{\leftrightarrow}{\partial}(t)\varphi^{\pm}_{n',p^{2'},p^{3'}}(x)
=\mp\delta_{nn'}\delta(p^{2}-p^{2'})\delta(p^{3}-p^{3'}).$$ One
can obtain the KE\index{Kinetic equation} (\ref{eq17}) by analogy
with the above considered case, assuming $r=0$,
\begin{eqnarray}\label{eqs5}
S_{n}(,P,t)=\frac{1}{2} w_{n}(P,t) \times \nonumber \\
\times \int\limits^{t}_{-\infty}dt' w_{n}((P,t'),t')
[1+2f_{n}(P,t')]\cos\left\{2\int_{t'}^{t}d\tau\omega_{n}(P,\tau)\right\}\,,
\end{eqnarray}
where  $\omega^2_{n}(P,t,t')=m^{2}+P^{2}(t,t')+|e|H(2n+1)$ and
\begin{equation}\label{eqs6}
w_{n}(P,t)=\frac{|e|E(t)P(t)}{\omega_{n}(P,t)}\,.
\end{equation}
\section{Back reaction problem}
For the description of the back reaction problem\index{back
reaction} it is necessary to add the regularized Maxwell equation.
It follows from Eq. (\ref{eq19})  that $w \sim P^{-2}$ at $P
\rightarrow \infty$. It leads to the absence of ultraviolet
divergences in the densities of observed quantities, which are
expressed by means of integrals containing the function
$f_{nr}(P,t)$. However, there is divergence of the sum over levels
$n$ and corresponding counter terms should subtracted while
calculating densities of physical values.

In order to construct counter-terms according to the procedure of
n-wave regularization \cite{Grib,1}, it is necessary to expand
functions $f$,$v$ and $z$ in asymptotic series under inverse
powers of one particle energy $\omega_{\alpha}$. After application
of this procedure to the set of Eqs. (\ref{eq24}) we obtain
leading contributions
$$ v^{\alpha}_3(t)=\frac{e \dot E(t)P}{4\omega_{\alpha}(P,t)}
\bigg(\frac{\epsilon_{\alpha}} {P}\bigg)^{g_{\alpha}-1} \,, \qquad
f^{\alpha}_4(t)=\left( \frac{eE(t)P}{4\omega_{\alpha}(P,t)}
\bigg(\frac{\epsilon_{\alpha}} {P}\bigg)^{g_{\alpha}-1}\right)^2
\,, $$ where $g_{\alpha}$ is the degeneracy factor for
corresponding set $\alpha$ of quantum numbers $\{n,r\}$ or
$\{n\}$.

The counter term is similar to the case $H=0$~\cite{EPJC} and sum
diverges logarithmically at $n \rightarrow \infty$ for fermions and
at $|P|\rightarrow\infty$ for scalar particles. Thus, we find that
the mean number density of $e^+e^-$
\index{Electron-positron pair production} pairs
$$n(t) = \sum _{nr}(P) f_{nr}(P,t)$$ is finite, whereas the mean energy
density and the total current density as the sum of the conduction
and the vacuum polarization currents contain divergences, which
can be eliminated with the help of counter terms:
 \begin{equation}\label{eq27}
\varepsilon(t) = 2\sum _{nr}(P) \left\{
\omega_{nr}(P)(f_{nr}(P,t)-f_c(P,t)) \right\}\,,
\end{equation}

\begin{equation}\label{eq28}
j^{-}(t) =2e \sum _{nr}(P) \frac{1}{\omega_{nr}(P)} \left\{ Pf_{nr}(P,t) +\frac{1}{2} \varepsilon_{nr}
 \left(v_{nr}(P,t)-v_c(P,t)\right)\right\}\,,
\end{equation}
 Expressions (\ref{eq27}) and (\ref{eq28}) are renormalized
according to charge renormalization procedure~\cite{1,EPJC}.
Here we have the short notation
\begin{equation}\label{eq29}
  \sum_{nr}(P) = \frac{eH}{(2\pi)^2} \sum_{r=1}^{g_{\alpha}}
  \sum_{n=0}^{\infty} \int_{-\infty}^{\infty} dP\,.
\end{equation}
The factor $eH$ describes the degeneracy of the distribution
function (\ref{eq16}) relatively of $p_{\bot}$ \cite{Land}.

The Eqs. system (\ref{eq24}) with the Maxwell Eq. $\dot{E} = -j^{\alpha}(t)$  compose the
complete equation system  of back reaction problem\index{back
reaction}.
\section{Conclusion and numerical results}
The creation of scalar particle in the time dependent electric
field with the presence of the strong collinear magnetic
field\index{Strong electromagnetic fields} is accompanied by the
increase of effective mass $m\rightarrow m^{*}_{n}=\left(m^{2}+
|e|H(2n+1)\right)^{1/2}$, so that we have even in the basic state
$m^{*}_{n=0}>m$. This circumstance decreases the vacuum production
of scalar bosons in the magnetic field \cite{Grib}.
\begin{figure}[ht]
 \centerline{ \epsfxsize=9cm\epsfbox{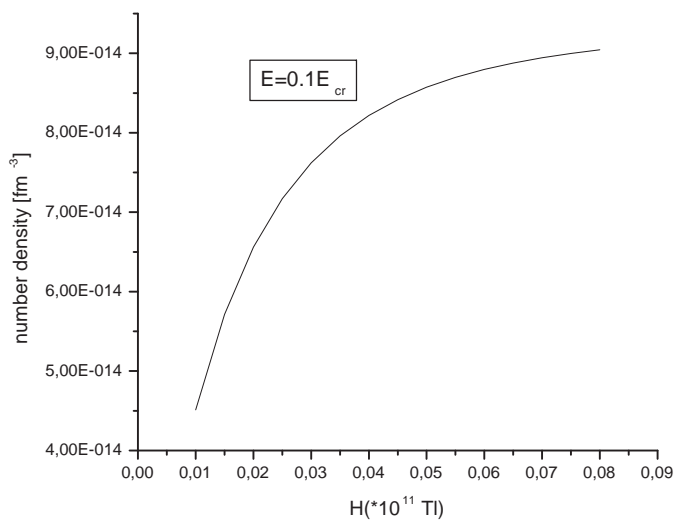}}
 \caption{Number density of fermions for $\vec E=(0,0,E_{0}cosh^{-2}(t/b))$, $b=0.5$
  as a function of magnetic field strength\label{fig1}.}
\end{figure}
\begin{figure}[ht]
 \centerline{\epsfxsize=8cm\epsfbox{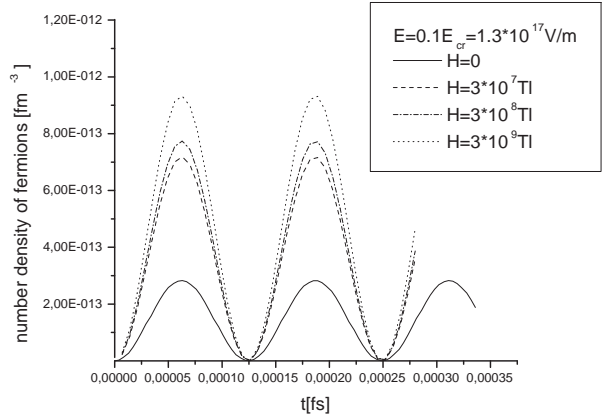}}
 \caption{Time evolution of fermionic number density for  $\vec E=(0,0,E_{0}sin(\Omega t))$,
 with $\Omega~=~8~\cdot~10^{18}\pi\ s^{-1}$, $\vec H=(0,0,H)$\label{fig2}.}
\end{figure}
Another situation is in the fermionic case: as it follows from
(\ref{eq19}), the effective mass coincides with the particle mass
for the spin states of electrons oriented opposite to the magnetic
field. That leads to paramagnetism of $e^+e^-$
plasma\index{Electron-positron plasma}, created from vacuum by the
EM field (\ref{eq1}). When spin oriented along the
magnetic field direction then the effective mass
$m_{2}^{*}=(m^{2}+2|e|Hn)^{1/2}$ and vacuum creation of such
electrons and positrons are suppressed. However, the creation of
the electrons in the state with $r=1$ with effective mass $m^{*}_{1}=m$ turns out
more intensive in comparison with the case of an absence of
magnetic field (this fact is well known for constant EM fields \cite{Grib}).
In Fig.~\ref{fig1}  we show the number density of $e^+e^-$ pairs after
action of external electric field impulse as a function of magnetic field strength.
In Fig.~\ref{fig2} we show the time evolution of number density in alternating external
electric field at various magnetic field strength.
\section*{Acknowledgments}
This work was supported by Deutsche Forschungsgemeinschaft (DFG)
under project number SCHM 1342/3-1 and RUS/17/102/00; and partly
by the  Education Ministry of  Russian Federation under grant
N~E00-33-20. We thank Prof.~A.~V.Prozorkevich for the helpful
discussion.

\printindex
\end{document}